# PHOTON ECHO QUANTUM MEMORY
# FOR ARBITRARY NON-STATIONARY LIGHT FIELDS

S.A.Moiseev[1], C. Simon[2] and N.Gisin[2]

[1] Kazan Physical-technical Institute of Russian Academy of Sciences, Kazan, Russia;
[2] Group of Applied Physics-Optique, University of Geneva, Switzerland

We develop the theory of an optical quantum memory protocol based on the three pulse *photon echo* (PE) in an optically dense medium with *controlled reversible inhomogeneous broadening* (CRIB). The wave-function of the retrieved photon echo field is derived explicitly as a function of an arbitrary input Data light field. The storage and retrieval of time-bin qubit states based on the described quantum memory is discussed, and it is shown that the memory allows to measure the path length difference in an imbalanced interferometer using short light pulses.

**PACS:** 03.67.Hk, 42.50.Ct, 42.50.Md.

### 1. Inroduction

The young field of Quantum Information (QI) Science already had significant impacts on physics, computer science and cryptology [Quantum Information Processing and Communication in Europe, Collection of 33 papers by the Europeans leaders, published by the European Communities, 2005 (ISBN 92-8924-3)]. It is truly remarkable how this cross-disciplinary activity has thrown new light on all those different fields. Quantum physicists learned information theory and realized how concepts from information theory illuminate their own field. Computer scientists had to revise some basic concepts and to realize that "information is physical". And cryptologists discovered that, on the one hand their favourite public key crypto-systems are vulnerable to quantum computers and, on the other hand, that quantum physics offers radically new crypto-systems [1].

In the last 10 years QI Science flourished because of these conceptual breakthroughs, but also because it was already offering some concrete examples of potential applications, like Quantum Key Distribution (QKD)[*] and Quantum Random Number Generators (QRNG). In particular, quantum optics greatly profited from and contributed to the development of Quantum Communication. QKD using flying photons was demonstrated and simultaneously beautiful experiments with entangled photons were performed.

---

[*] Let us emphasize that QKD allows one not only to encrypt confidential messages, but also other cryptographic applications, like broadcasting through 1-to-1 channels.

Today there are research groups on QKD in a large number of countries. It has been well understood what the needs and limitations of QKD are (at least from a physicist's point of view, a lot still has to be done on the side of the cryptography community!). In particular, it is well understood that single-photons and entangled photons are not needed, coherent states of light suffice\footnote{Either of mesoscopic size – tens of photons per pulse - called continuous variable QKD with homodyne detection, or of size smaller than 1 photon, called weak pulse QKD. The first case imposes huge demands on the classical post-processing to correct for the unavoidable errors due to quantum fluctuations, the second case allows one to post-select with photon counters the non-empty pulses, a trick that much simplifies the classical post-processing.}. It is also well understood that the small but non-zero losses of telecom optical fibers together with the dark counts of photon-detectors limit the maximal distance over which QKD is possible (depending on the detector, the limit ranges from tens of km to possibly a few hundreds). Accordingly, QKD is in part leaving QI Science to enter QI Engineering.

QKD is no longer a potential application, but is already a present one! The future of Quantum Communication and the future role of quantum optics for research in QI Science have to be found in new challenges. A natural and promising candidate is a quantum repeater for flying telecom photons [2]. This is a really big challenge with huge potential applications. Indeed, ideal quantum repeaters would allow one to arbitrarily extend the distances and the complexity of QKD networks. To our present day knowledge, this would require quantum memories, entangled photons and entanglement swapping, all beautiful processes already partially demonstrated, but that are still very far from real life applications (for instance, none of the few demonstrations use compatible wavelengths). This big challenge will require lots of research by quantum opticians, material scientists and system integration specialists: good work for the next decade!

In this contribution, we concentrate on the quantum memory part [3-10] of the grand picture sketched above. Indeed, today this is the weakest part. The ultimate goal is a quantum memory (QM) able to store for milliseconds or longer and to retrieve on demand arbitrary light pulses of ns durations, hence with large spectra, at a wavelength compatible with telecom optical fibers. This is a long term objective; hence there is also a need for intermediate goals. Recently some of us proposed a protocol inspired by the 3-pulse photon echo technique with *controlled reversible inhomogeneous broadening* (CRIB) [6] that – in principle – allows one to realize a complete quantum memory for photons, fully compatible with the 1-dimensional guiding structure of single-mode optical fibers [9]. In the next section we recall this proposal, describe it in the Schrödinger picture and provide the explicit out-coming wave-function, in function of the in-coming one. Then we discuss the storage and retrieval of time-bin qubit states and demonstrate how the quantum memory can be used for the direct measurement of the path-length difference *ΔL* in an imbalanced interferometer using pulses much shorter than *ΔL/c* .

## 2. Quantum memory based on the three pulse photon echo with controlled reversible inhomogeneous broadening

The analyzed QM protocol is based on using the traditional properties of the photon echo [11] for the storage of optical information in the spectral profile of inhomogeneously broadened atomic excitations [12,13,14]. The specific requirements of a QM based on photon echo with CRIB [6-9] are high optical density of the resonant atomic system and complete reversibility of the quantum dynamics in the interacting quantum light fields and atomic system. Recent experiments on primary two pulse photon echo in the optically dense media have demonstrated a large amplitude of the echo signal pulse which was comparable with the amplitude of the initial Data light field [15]. However the large amplitude was reached due to the amplification of the generated photon echo, which can not be used for perfect reconstruction of arbitrary quantum states of the Data light field. Other spatial schemes in gaseous media which were similar to the QM photon echo with CRIB have been studied earlier on in experiments on three-level photon echo [16]. We also note a recent experiment [17] demonstrating strong superfluorescent properties in the photon echo emission, which are important for the high efficiency of the QM based on the photon echo.

First experiment towards the realization of QM photon echo with CRIB were recently performed in an optically thin medium [10] using a classical Data light field. The authors of [10] have successfully realized the CRIB of the resonant atomic transition, which is one of the two important key procedures of the analyzed QM (see below). In comparison with well known QM based on the electromagnetically induced transparency (EIT) technique [5] which uses a near adiabatic slow light-atomic system interaction, the QM photon-echo utilizes a fast complete absorption of the Data light pulse on the resonant transition. Despite this difference there are some common properties in both techniques, such as the use of long-lived atomic levels for storage of the quantum light states, where the QM photon echo becomes especially close to fast limit of QM based on EIT [18].

We note that the analyzed optical quantum memory is based on the non-adiabatic interactions of light with resonant atomic medium, which can be realized in principle for very short and intense pulses of light where nonlinear effects in the light-medium interaction can play a significant role in absorption and emission of the quantum light fields. Basic scheme of storage and retrieval the quantum light state is depicted in the Fig.1. The scheme represents a specific time delay variant of backward four wave mixing processes where the direction of photon echo emission is determined by the following phase matching condition $k_{echo} = k_- = k_+ - K_1 + K_2$ (where $k_{+,-}$ are the wave vectors of Data (forward) and backward (echo) quantum fields, $K_{1,2}$ are the wave vectors of first and second control laser fields). Total Hamiltonian is the following:

$$\hat{H} = \hat{H}_{ph} + \hat{H}_a(t) + \hat{V}_{a-ph} + \hat{V}_{cont}(t), \qquad (1)$$

where $\hat{H}_{ph}$, $\hat{H}_a$ and $\hat{V}_{a-ph}$ are the energies of quantum fields $E_{+,-}$, atoms and their interaction; $\hat{V}_{coh}(t) = \hat{V}_1(t) + \hat{V}_2(t)$ is the interaction of atoms with two short intense control laser pulses treated classically with given parameters of their electric fields. We suggest that the two control laser pulses have a short temporal duration $\delta t$ and so can be taken into account only near to the fixed moments of time $t_1$ and $t_2$ (see Fig.1 and Fig.2). We represent Hamiltonian of the atomic system $H_a$ as a sum of two parts:

$$\hat{H}_a(t) = \hat{H}_a^o + \delta\hat{H}_a(t), \qquad (2a)$$

$$\hat{H}_a^o = \hbar\sum_{j=1}^{N}\{\omega_{31}\hat{P}_{33}^j + \omega_{21}\hat{P}_{22}^j\}, \qquad (2b)$$

$$\delta\hat{H}_a(t) = \hbar\sum_{j=1}^{N}\Delta_j(t)\hat{P}_{33}^j, \qquad (2b)$$

where $\delta\hat{H}_a(t)$ is determined by inhomogeneous broadening of the resonant frequencies on the atomic transitions 1-3 and 2-3, $\hat{P}_{mn}^j = |m\rangle_{jj}\langle n|$ is an atomic operator of j-th atom, $\hbar$ is the Planck's constant.

Hamiltonian $\hat{H}_{ph}$ of the two quantum light fields $E_+$ and $E_-$ is:

$$\hat{H}_{ph} = \hbar\int dk\,\omega_k \hat{a}_k^+ \hat{a}_k = c\hbar\sum_{\sigma=+,-}\int d\kappa \cdot |\kappa + \sigma k_o|\hat{a}_{\kappa+\sigma k_o}^+ \hat{a}_{\kappa+\sigma k_o} = \hat{H}_{ph}^o + \sum_{\sigma=+,-}\delta\hat{H}_{ph,\sigma}, \qquad (3a)$$

$$\hat{H}_{ph}^o = \hbar\omega_{31}\sum_{\sigma=+,-}\int d\kappa\,\hat{a}_{\kappa+\sigma k_o}^+ \hat{a}_{\kappa+\sigma k_o}, \qquad (3b)$$

$$\delta\hat{H}_{ph,\sigma} = \sigma c\hbar\int d\kappa\cdot \kappa\,\hat{a}_{\kappa+\sigma k_o}^+ \hat{a}_{\kappa+\sigma k_o} = -i\sigma c\hbar\int_{-\infty}^{\infty}dz\,\hat{A}_\sigma(z)\tfrac{\partial}{\partial z}\hat{A}_\sigma^+(z), \qquad (3c)$$

where $\sigma = +,-$ stay for forward and backward propagating light beams, it has been assumed $ck_o=\omega_o=\omega_{31}$ ($\omega_{31}$ is resonant frequency on the atomic transition $|1\rangle \leftrightarrow |3\rangle$).

We introduce the quantum field operators $E_\sigma^+(t,z) = i(\frac{\hbar\omega_o}{2\varepsilon_o S})^{1/2}\exp\{-i(\omega_o t - \sigma k_o z)\}\hat{A}_\sigma(z)$ assuming that effective one-dimensionality of the system is ensured either by a large Fresnel number $F=S/\lambda_o L\gg 1$ (where $\lambda_o=2\pi/k_o$ is carrier wavelength of the quantum light, $L$ – length of the medium, S is a cross section of light beams, $\varepsilon_o$ is the electric permittivity [19]), or by a waveguide structure, and sufficiently narrow bandwidth of the quantum light field $\Delta\omega \approx c|\kappa_{max}|<ck_o$, where

$$\hat{A}_\sigma(z) = (2\pi)^{-1/2} \int d\kappa \hat{a}_{\kappa+\sigma k_o} e^{i\kappa z} \qquad (4)$$

are slowly varied operators of the forward and backward fields.

We evaluate the dynamic Schrödinger equation transferring to the new representation $|\Psi\rangle = \hat{U}(t-t_o)|\tilde{\Psi}(t)\rangle$ (where $\hat{U}(t-t_o) = \exp\{-i(t-t_o)(\hat{H}_a^o + \hat{H}_{ph}^o)\}$)

$$i\frac{d}{dt}|\tilde{\Psi}\rangle = \hat{H}'|\tilde{\Psi}\rangle, \qquad (5a)$$

with Hamiltonian

$$\hat{H}'(t) = \delta\hat{H}_a(t) + \sum_{\sigma=+,-}\{\delta\hat{H}_{ph,\sigma} + \hat{V}'_{a-ph}(\hat{A}_\sigma, \hat{S}_{31,\sigma})\} + \hat{V}'_{contr}(t), \qquad (5b)$$

$$\hat{V}'_{a-ph}(\hat{A}_\sigma, \hat{S}_{31,\sigma}) = -\hbar g \sum_{j=1}^N \hat{A}_\sigma(z_j)\hat{S}_{31;\sigma}^j + H.C., \qquad (5c)$$

where $\hat{V}'_q = \hat{U}^+(t-t_o)\hat{V}_q\hat{U}(t-t_o)$, $\hat{S}_{31;\sigma}^j = \hat{P}_{31}^j e^{i\sigma k_o z_j}$; we have taken into account narrow light bandwidth $\Delta\omega \approx c|\kappa_{max}| \ll ck_o$ so that the coupling constant g depends negligibly weak on the light frequency near to the atomic frequency $\omega_{31}$: $g(\omega) \cong g = id(\frac{\omega_o}{2\hbar\varepsilon_o S})^{1/2}$ (d is an dipole moment on the atomic transition $|1\rangle \leftrightarrow |3\rangle$).

Before analyzing Eq. (5), we assume that the frequency detuning $\Delta_j(t)$ of j-th atom is controlled in time in accordance with the following relation

$$\Delta_j(t = 2t_{inv} - t') = -\Delta_j(t' < t_1), \qquad (6a)$$

$$\delta\hat{H}_a(2t_{inv} - t') = -\delta\hat{H}_a(t') = -\delta\hat{H}_a(2t_{inv} - t). \qquad (6b)$$

Spectral inversion of the atomic detunings at the time of $t = t_{inv}$ on the atomic transition 1-3 in Eq. (6) should exist for the interaction of atoms with the light fields $E_+$ and $E_-$ (for the times $t' < t_1$ and $t > t_2$, see Fig.2). The spectral inversion of inhomogeneous broadening provides perfect rephasing of the atomic coherence for $t > t_2$ leading thereby to the photon echo signal emission similar to spin echo effect in nuclear magnetic resonance with switching of the magnetic field orientation. Such spectral inversion of atomic detuning occurs for the interaction of atoms with the forward and backward light beams in gaseous medium due to the opposite sign of Doppler shift. Similar spectral inversion can be performed using a hyperfine interaction control [7] and artificial switching of the external electric (magnetic) field gradients [8-10]. In contrast to the nuclei spin echo, the analyzed photon echo can provide a perfect reconstruction of the initial quantum state due to the large optical density of the resonant transition $|1\rangle \leftrightarrow |3\rangle$ (see below)

The initial quantum state at the time of $t = t_o$ is:

$$|\tilde{\Psi}(t_o)\rangle = |\tilde{\psi}_+(\hat{a}_{\sigma=+}^+, t_o)\rangle_{ph} \otimes |1\rangle_a, \qquad (7a)$$

where $|1\rangle_a = \prod_{j=1}^{N} \otimes |1_j\rangle$ is a ground state of the atomic medium and the state $|\tilde{\psi}_+(\hat{a}^+_{\sigma=+}, t_o)\rangle_{ph}$ describes the Data quantum light pulse propagating along the forward z-direction:

$$|\tilde{\psi}_+(\hat{a}^+_{\sigma=+}, t_o)\rangle_{ph} = \{\varphi_o + \int d\kappa \varphi_1(\kappa)\hat{a}^+_{\kappa+k_o} + \tfrac{1}{\sqrt{2!}}\iint d\kappa_2 d\kappa_1 \varphi_2(\kappa_2, \kappa_1)\hat{a}^+_{\kappa_2+k_o}\hat{a}^+_{\kappa_1+k_o} + ...\}|0\rangle_{ph}$$

$$= \{\varphi_o + \sum_{n=1}^{N_{max}} \tfrac{1}{\sqrt{n!}} \int d\kappa_n ... \int d\kappa_1 \varphi_n(\kappa_n, ..., \kappa_1)\hat{a}^+_{\kappa_n+k_o} \cdot ... \cdot \hat{a}^+_{\kappa_1+k_o}\}|0\rangle_{ph}$$

$$= \{\varphi_o + \sum_{n=1}^{N_{max}} \tfrac{1}{\sqrt{n!}} \int dz_n ... \int dz_1 \tilde{\varphi}_n(z_n, ..., z_1)\hat{A}^+_+(z_n) \cdot ... \cdot \hat{A}^+_+(z_1)\}|0\rangle_{ph}. \quad (7b)$$

where for simplicity we have ignored the polarization degree of freedom of the light fields assuming $\tilde{\varphi}_n(z_n, ..., z_1) = (2\pi)^{-n/2} \int d\kappa_n ... \int d\kappa_1 \varphi_n(\kappa_n, ..., \kappa_1) e^{-i(\kappa_n z_n + ... + \kappa_1 z_1)}$ with normalization

$$|\varphi_o|^2 + \sum_{n=1}^{N_{max}} \int d\kappa_n ... \int d\kappa_1 |\varphi_n(\kappa_n, ..., \kappa_1)|^2 = |\varphi_o|^2 + \sum_{n=1}^{N_{max}} \int dz_n ... \int dz_1 |\tilde{\varphi}_n(z_n, ..., z_1)|^2 = 1.$$

The polarization properties play often an important role in photon echo experiments [20] and they should be taken into account for practical realizations of the QM photon echo protocol. Using Eqs. (5a) and (7) we write the wave function $|\tilde{\Psi}(t > t_o)\rangle$ before entrance of the Data light pulse to the medium at the time of $t \approx 0$

$$|\tilde{\Psi}(t)\rangle = \exp\{-ic(t-t_o)\int dk \cdot k \hat{a}^+_{\kappa+k_o} \hat{a}_{\kappa+k_o}\}|\tilde{\Psi}(t_o)\rangle$$

$$= \{\varphi_o + \sum_{n=1}^{N_{max}} \tfrac{1}{\sqrt{n!}} \int d\kappa_n ... \int d\kappa_1 \exp[-ic(t-t_o)\sum_{m=1}^{n} \kappa_m]\varphi_n(\kappa_n, ..., \kappa_1)\hat{a}^+_{\kappa_n+k_o} \cdot ... \cdot \hat{a}^+_{\kappa_1+k_o}\}|0\rangle_{ph}, \quad . \quad (7c)$$

Below we consider an arbitrary dependence of the wave function $\varphi_n(\kappa_n, ..., \kappa_1)$ on $\kappa_n, ..., \kappa_1$, which can describe entangled states of the multi-photon "time-bin" light fields used for the quantum communication (see [1]). For convenience we choose phase factors in the function $\varphi_n(\kappa_n, ..., \kappa_1)$ which correspond to the appearance of the Data pulse in the medium near to t=0 and z=0 (see Fig.1.). In particular case of a Gaussian single photon wave packet, we have $\varphi_1(\kappa_1) = (\sqrt{2\pi}\delta k)^{-1/2} \exp\{-\tfrac{1}{2}(\kappa_1/\delta k)^2\}e^{i\phi_1}$ and a single-photon state where the photon is in a superposition of being in either of two wave packets is:

$$\varphi_1(\kappa_1, \tau) = (\sqrt{2\pi}\delta k)^{-1/2} \exp\{-\tfrac{1}{2}(\kappa_1/\delta k)^2\}\{\alpha e^{i\phi_1} + \beta e^{i(\phi_2 + c\kappa_1 \tau)}\}, \quad (8)$$

where $\tau$ is a time delay of the second wave packet in comparison to the first packet, $\alpha$ and $\beta$ are the wave packet amplitudes ($|\varphi_o|^2 + |\alpha|^2 + |\beta|^2 = 1$ for $\tau^{-1} << c\delta k$), $\phi_{1,2}$ are constant phase shifts of the packets; $\delta k = \delta\omega/c$, $\delta\omega$ is a spectral width of the light field. The n-th photon functions $\varphi_n(\kappa_n, ..., \kappa_1)$ with many wave packets can be presented similarly.

The coherent interaction of the Data field with atomic system evolves without generation of backward scattering field in the spatially homogeneous medium. Therefore for $t<t_1$ the Hamiltonian of Eq. (5a) $\hat{H}'$ is reduced to

$$\hat{H}'(t<t_2) = \hat{H}'(\hat{S}_{31,+}, \hat{A}_+,...;t) = \delta\hat{H}_a(t) - ic\hbar\int_{-\infty}^{\infty} dz \hat{A}_+(z) \tfrac{\partial}{\partial z} \hat{A}_+^+(z) + \hat{V}'_{a-ph}(\hat{A}_+, \hat{S}_{31,+}), \qquad (9)$$

where $\delta\hat{H}_a(t<t_1) = \hbar\sum_{j=1}^{N} \Delta_j(t)\hat{S}_{31,+}^j \hat{S}_{13,+}^j$.

Note that the Hamiltonian $\hat{H}'_+(t)$ is invariant to the global gauge transformation of atomic and field's operators: $\hat{S}_{31;+}^j \rightarrow e^{i\vartheta}\hat{\tilde{S}}_{31;+}^j$, $\hat{A}_+(z) \rightarrow e^{-i\vartheta}\hat{\tilde{A}}_+(z)$ (where $\hat{\tilde{S}}_{31;+}^j$ and $\hat{\tilde{A}}_+(z)$ are time-independent operators, which have the same commutation relations). It is interesting that the gauge transformation plays an important role in the quantum dynamics analysis of the echo signal irradiation. Using Hamiltonian in Eq. (9) we write the solution of Eq.(5) in the following general form:

$$\left|\tilde{\Psi}(t)\right\rangle = \hat{T}_1(\hat{S},\hat{A};t,t_o)\left|\tilde{\Psi}(t_o)\right\rangle, \qquad (10a)$$

$$\hat{T}_1(\hat{S},\hat{A};t,t_o) = \hat{T}\exp\{-i\hbar^{-1}\int_{t_o}^{t} dt'\hat{H}'(\hat{S}_{31,+},\hat{A}_+,...;t')\}, \qquad (10b)$$

where $\hat{T}$ denotes the time-ordered product,

$$\hat{T}_1(\hat{S},\hat{A};t,t_o) = Lim_{M\rightarrow\infty} \exp\{-i\hbar^{-1}\Delta t\hat{H}'(t_M)\}\cdot...\cdot\exp\{-i\hbar^{-1}\Delta t\hat{H}'(t_1)\}\exp\{-i\hbar^{-1}\Delta t\hat{H}'(t_o)\}. \qquad (10c)$$

Here $\Delta t = (t-t_o)/M$, $t_m = t_o + m\Delta t = t$, $t_M = t$.

The wave function $\left|\tilde{\Psi}(t)\right\rangle$ can be found analytically [21] for arbitrary initial quantum states in Eq. (7) in the simplest case of constant atomic detunings $\Delta_j(t) = \Delta_j$. The solution demonstrates an irreversible absorption of the Data light pulse by the optically dense medium due to the complete dephasing of the atomic polarization in the presence of the inhomogeneously broadened atomic transition. Therefore eventually the state $\left|\tilde{\Psi}(t)\right\rangle$ in Eq.(10) evolves to the separable state of the atoms and light field:

$$\left|\tilde{\Psi}(t)\right\rangle \cong \left|0\right\rangle_{ph}\left|\phi(\beta(t),\hat{S}_{31,+})\right\rangle_a, \qquad (11a)$$

where the atomic wave function is formularized through the decomposition on atomic operators $S_{31,+}^j$ (j=1,2,...,N):

$$\left|\phi(\beta(t),e^{i\gamma}\hat{S}_{31,+})\right\rangle_a = \{\varphi_o + \sum_{j_1}\beta(j_1,t)e^{i\gamma_1}\hat{S}_{31,+}^{j_1} + \sum_{j_2,j_1}\beta(j_2,j_1;t)e^{i\gamma_2}e^{i\gamma_1}\hat{S}_{31,+}^{j_2}\hat{S}_{31,+}^{j_1} + ...\}\left|1\right\rangle_a$$

$$= \{\varphi_o + \sum_{n=1}^{N_{max}}\sum_{j_2,j_1}\beta(j_n,...,j_1;t)e^{i\gamma_{j_n}}\hat{S}_{31,+}^{j_n},...,e^{i\gamma_1}\hat{S}_{31,+}^{j_1}\}\left|1\right\rangle_a. \qquad (11b)$$

The atomic state of Eq. (11b) is excited only if the energy of Data light pulse is insufficient to produce non-absorbing solitons. Functions $\beta(j_n,...,j_1;t)$ characterize the state with n excited atoms, phase factor $\gamma_{j_n}$ characterizes a further atomic evolution. Temporal behavior of the atomic wave functions $\beta(j_n,...,j_1;t) = \exp\{-i\sum_{m=1}^{n}\Delta_{j_m}t\}\beta_o(j_n,...,j_1)$ occurs for the vacuum field state. We note the function $\beta_o(j_n,...,j_1)$ corresponds unambiguously to the Data field function $\varphi_n(\kappa_n,...,\kappa_1)$ (and $\tilde{\varphi}_n(z_n;...;z_1)$). Using of the solution $|\tilde{\Psi}(t)\rangle$ in the general form of Eq. (11b) will be enough for our further analysis.

We note that the atomic coherence in Eq. (11b) can be transferred to the long-lived atomic coherence between the levels 1 and 2 by the first short control laser pulse coupling the states $|3\rangle$ and $|2\rangle$

$$|\tilde{\Psi}(t_1+\delta t)\rangle = \hat{U}_1|\tilde{\Psi}(t_1)\rangle = |0\rangle_{ph}|\phi(\beta(t_1),e^{-i\chi_1}\hat{S}_{21})\rangle_a, \quad (12)$$

where $\hat{U}_l = \prod_{j=1}^{N}\hat{U}_l^j$, (l=1,2), $\hat{U}_l^j = \exp\{-(i/\hbar)\int \hat{V}_l'(t')dt'\}\Big|_{\vartheta_l=\int dt\Omega_l(t')=\pi}$

$= P_{11}^j + i\{P_{23}^j\exp(-i\chi_{1,j}) + h.c.\}$, $\vartheta_l$ and $\Omega_l$ are the pulse area and Rabi frequency of the *l*-th control laser pulse, $\chi_{l,j} = \omega_{32}(t_j - (-1)^l z_j/c) + \xi_j$, $\xi_{1,2}$ are the phases of the first and second laser pulses.

After the first laser pulse impact (t > $t_1+\delta t$) we switch on the atomic detuning (near to time t=$t_2$ in the Fig.2). For simplicity we assume that the stored quantum state of the Data field is not destroyed on the atomic transition 1-2 during the time interval $t_1 < t < t_2$ due to the weak decoherent processes. Then the second control laser pulse with the pulse area $\vartheta_2=\pi$ is applied at the moment of time $t = t_2 = 2t_{inv} - t_1$ (see Fig.2). We obtain the following wave function after the second laser pulse (t>$t_3$)

$$|\tilde{\Psi}(t_3+\delta t)\rangle \cong \hat{U}_2|\tilde{\Psi}(t_3)\rangle = |0\rangle_{ph}|\phi(\beta(t_1)e^{-i\mu}\hat{S}_{31,+})\rangle_a\Big|_{2\omega_{21}L/c<<1} \cong |0\rangle_{ph}|\phi_r(\beta(t_1)\hat{R}_{31,-})\rangle_a, \quad (13)$$

where we have introduced the new atomic operators $\hat{R}_{31,-}^j = \hat{P}_{31}^j e^{-i(k_o z_j + \chi_{12} + \pi)} = \hat{S}_{31,-}^j e^{-i(\chi_{12}+\pi)}$, $\mu_{j_n} = (2\omega_{32}z_{j_n}/c + \chi_{12} + \pi)$ is a new atomic phase, $\chi_{12} = \xi_1 - \xi_2 - \omega_{32}t_{21}$ [6, 22].

Atomic operators $\hat{S}_{31,-}$ and $\hat{R}_{31,-}$ in state of Eq. (13) correspond to polariton waves propagating along the –z direction so the atomic system can irradiate coherently only the light fields in the same direction. We consider a small enough splitting between the two lowest atomic levels $\omega_{21}/(c\alpha) << \pi$ where α is the absorption coefficient on the transition 1-3 therefore we can approximately assume $\omega_{32}z/c \approx k_o z$ for the atoms participating in the interactions with quantum fields. The phase matching conditions can be accurately fulfilled in the ladder three level scheme of the atomic transition [22].

Below we analyze a reconstruction of the initial Data light field state in Eq. (7) at unitary evolution of the interacting quantum light and atomic system. We rewrite the Hamiltonian (5a) taking into account only the interaction of the atoms with backward modes $E_-$ transferring to the new field operators $\hat{\tilde{A}}_-(z_j) = e^{i\chi_{12}}\hat{A}_-(z_j)$:

$$\hat{V}'_{a-ph}(\hat{A}_-, \hat{S}_{31,-}) = -\hbar g \sum_{j=1}^{N} \hat{A}_-(z_j)\hat{S}^j_{31;-} + H.C. = \hbar g \sum_{j=1}^{N} \hat{\tilde{A}}_-(z_j)\hat{R}^j_{31;-} + H.C. = -\hat{V}'_{a-ph}(\hat{\tilde{A}}_-, \hat{R}_{31,-}). \quad (14)$$

Thus the Hamiltonian of Eq. (5b) gets the form:

$$\hat{H}'(t \geq t_2) = -\{\delta\hat{H}_a(2t_{inv} - t) - ic\hbar \int_{-\infty}^{\infty} dz \hat{\tilde{A}}_-(z)\frac{\partial}{\partial z}\hat{\tilde{A}}_-^+(z) + \hat{V}'_{a-ph}(\hat{\tilde{A}}_-, \hat{R}_{31,-})\}$$

$$= -\hat{H}'(\hat{\tilde{A}}_-, \hat{R}_{31,-}, 2t_{inv} - t) \quad (15)$$

where $\delta\hat{H}_a(2t_{inv} - t) = \hbar \sum_{j=1}^{N} \Delta_j(2t_{inv} - t)\hat{R}^j_{31,-}\hat{R}^j_{13,-}$.

We note that there are identical commutation relations in the two groups of atomic and field's operators: { $\hat{R}^{j_n}_{31,-}$ and $\hat{\tilde{A}}_-(z_n)$ in Eqs. (14)-(15)} and { $\hat{S}^j_{31;+}$ and $\hat{A}_+(z_j)$ in Eq. (10)}. Therefore the algebraic structure of Hamiltonian $\hat{H}'(\hat{\tilde{A}}_-, \hat{R}_{31,-}, 2t_{inv} - t)$ completely coincides with the structure of the initial Hamiltonian $\hat{H}'(\hat{S}_{31,+}, \hat{A}_+, ...; t)$ in Eq. (10). Thus using Eq. (15) we find the wave function $|\tilde{\Psi}(t > t_2)\rangle$ in the following form:

$$|\tilde{\Psi}(t)\rangle = \hat{T}_2(\hat{R}, \hat{\tilde{A}}; t, t_2)|\phi_r(\beta(t_1)\hat{R}_{31,-})\rangle_a. \quad (16a)$$

where

$$\hat{T}_2(\hat{R}, \hat{\tilde{A}}; t, t_2) = \hat{T}\exp\{i\hbar^{-1}\int_{t_2}^{t} dt''\hat{H}'(\hat{R}_{31,-}, \hat{\tilde{A}}_-, ...; 2t_{inv} - t'')\} \quad (16b).$$

We will show that the unitary evolution of $|\tilde{\Psi}(t > t_2)\rangle$ in Eq. (16) can evolve reversible in time to the pure field state and ground state of the atomic system. Using the identical commutation relations in the wave function of Eq. (16) we rewrite the state $|\phi_r(\beta(t_1)\hat{R}_{31,-})\rangle_a$ in the form of Eqs. (7) and Eq. (10) replacing the operators $\hat{S}^j_{31;+}$ and $\hat{A}_+(z_j)$ by the operators $\hat{R}^j_{31,-}$ and $\hat{\tilde{A}}_-^+(z_j)$:

$$|\phi_r(\beta(t_1)\hat{R}_{31,-})\rangle_a = \hat{T}_1(\hat{R}, \hat{\tilde{A}}; t, t_o)|\tilde{\psi}_-\rangle_{ph}|1\rangle_a, \quad (17a)$$

where we have introduced the state

$$|\tilde{\psi}_-\rangle_{ph}|1\rangle_a = \{\varphi_o + \sum_{n=1}^{N_{max}} \frac{1}{\sqrt{n!}}\int dz_n ... \int dz_1 \tilde{\varphi}_n(z_n,...,z_1)\hat{\tilde{A}}_-^+(z_n)\cdot...\cdot\hat{\tilde{A}}_-^+(z_1)\}|0\rangle_{ph}|1\rangle_a . \qquad (17b)$$

New equations. (17a) and (17b) differ from Eqs. (10) and (7) only by the substitutions $\hat{S}_{31;+}^j \leftrightarrow \hat{R}_{31;-}^j$ and $\hat{A}_+(z) \leftrightarrow \hat{\tilde{A}}_-(z)$. Therefore we can write the function $|\tilde{\Psi}(t)\rangle$ for $t \geq t' = t_2 + t_1 - t_o$ ($t_2 = 2t_{inv} - t_1$) (see Fig.2) in the following form:

$$|\tilde{\Psi}(t \geq t' = t_2 + t_1 - t_o)\rangle = \hat{T}_2(\hat{R},\hat{\tilde{A}};t,t')\hat{T}_2(\hat{R},\hat{\tilde{A}};t',t_2)\hat{T}_1(\hat{R},\hat{\tilde{A}};t_1,t_o)|\tilde{\psi}\rangle_{ph}|1\rangle_a . \qquad (18)$$

Using a decomposition of the exponential operator in Eq. (18) similar to Eq. (10c), we find

$$\hat{T}_2(\hat{R},\hat{\tilde{A}};t',t_2)\hat{T}_1(\hat{R},\hat{\tilde{A}};t_1 t_o) =$$

$$Lim_{M\to\infty} \exp\{i\hbar^{-1}\Delta t\hat{H}'(2t_{inv} - t_2 - M\Delta t)\}\cdot...\cdot\exp\{i\hbar^{-1}\Delta t\hat{H}'(2t_{inv} - t_2 - \Delta t)\}\exp\{i\hbar^{-1}\Delta t\hat{H}'(2t_{inv} - t_2)\}$$

$$\exp\{-i\hbar^{-1}\Delta t\hat{H}'(t_o + M\Delta t)\}\exp\{-i\hbar^{-1}\Delta t\hat{H}'(t_o + (M-1)\Delta t)\}\cdot...\cdot\exp\{-i\hbar^{-1}\Delta t\hat{H}'(t_o)\} = 1, \qquad (19)$$

where we have used Eqs. (5a), (9) and Eq. (15) and taken into account

$$\hat{H}'(2t_{inv} - t_2 = t_1) = \hat{H}'(t_o + M\Delta t = t_1), \quad \hat{H}'(2t_{inv} - t_2 - \Delta t) = \hat{H}'(t_o + (M-1)\Delta t) = \hat{H}'(t_1 - \Delta t), \quad ...,$$

$$\hat{H}'(2t_{inv} - t_2 - M\Delta t) = \hat{H}'(t_o)$$

Thus we obtain

$$|\tilde{\Psi}(t \geq t')\rangle = \hat{T}_2(\hat{R},\hat{\tilde{A}};t,t')|\tilde{\psi}(a_{\sigma=-}^+,t')\rangle_{ph}|1\rangle_a , \qquad (20a)$$

where

$$|\tilde{\psi}(a_{\sigma=-}^+,t')\rangle_{ph} = \{\varphi_o + \sum_{n=1}^{N_{max}} \frac{1}{\sqrt{n!}}\int dz_n ... \int dz_1 \exp\{-in\chi_{12}\}\tilde{\varphi}_n(z_n,...,z_1)\hat{A}_-^+(z_n)\cdot...\cdot\hat{A}_-^+(z_1)\}|0\rangle_{ph} .$$

$$= \{\varphi_o + \sum_{n=1}^{N_{max}} \frac{1}{\sqrt{n!}}\int d\kappa_n ... \int d\kappa_1 \exp\{-in\chi_{12}\}\varphi_n(\kappa_n,...,\kappa_1)a_{\kappa_n-k_o}^+ \cdot...\cdot a_{\kappa_1-k_o}^+\}|0\rangle_{ph} \qquad (20b)$$

Since the time $t_o \ll 0$ corresponds to the initial Data light pulse before its interaction with the atomic system, hence the operator $\hat{T}_2(\hat{R},\hat{\tilde{A}};t-t')$ describes a free propagation of the emitted light pulse-photon echo field situated far off the atomic system at $t \geq t'$. So taking into account the ground state of the atomic system in Eq. (20a) we obtain $\hat{H}'(t > t') = \delta\hat{H}_{ph,-}$ in Eq. (16b) which leads to following final state of $|\tilde{\Psi}(t \geq t')\rangle$:

$$|\tilde{\Psi}(t \geq t')\rangle$$

$$= \{\varphi_o + \sum_{n=1}^{N_{max}} \frac{1}{\sqrt{n!}}\int d\kappa_n ... \int d\kappa_1 \exp\{i[c(t-t')\sum_{m=1}^{n}\kappa_m - n\chi_{12}]\}\varphi_n(\kappa_n,...,\kappa_1)a_{\kappa_n-k_o}^+ \cdot...\cdot a_{\kappa_1-k_o}^+\}|0\rangle_{ph}|1\rangle_a$$

$$(21)$$

The final Eq. (21) gives the general analytical solution for an arbitrary quantum state of light reconstructed in the photon echo protocol under consideration. Thus the proposed technique represents an ideal quantum machine of reversible unitary reconstruction for arbitrary stored quantum states of light pulses. Applying the general solution in Eq. (21) to the initial single photon state with two wave packets Eq. (8) (adding a vacuum component $\varphi_o |0\rangle_{ph}$) we analyze the properties of the retrieved field state in comparison to the properties of the Data field state $|\tilde{\Psi}_1(t \geq t')\rangle$

$$= \{\varphi_o + (\sqrt{2\pi}\delta k)^{-1/2}\alpha\int d\kappa_1 \exp\{-\tfrac{1}{2}(\kappa_1/\delta k)^2\}\exp\{i[c(t-t')\kappa_1 + \phi_1 - \chi_{12}]\}a^+_{\kappa_1-k_o}\}|0\rangle_{ph}|1\rangle_a$$
$$+ (\sqrt{2\pi}\delta k)^{-1/2}\beta\int d\kappa_1 \exp\{-\tfrac{1}{2}(\kappa_1/\delta k)^2\}\exp\{i[c(t-t'+\tau)\kappa_1 + \phi_2 - \chi_{12}]a^+_{\kappa_1-k_o}\}|0\rangle_{ph}|1\rangle_a. \quad (22)$$

Using Eq. (22) and Eq. (4) we find the field amplitudes of the initial Data field $A_+(t<0,z) = \langle\tilde{\Psi}_1(t)|\hat{A}_+(z)|\tilde{\Psi}_1(t)\rangle$ and of the retrieval field $A_-(t>t',z) = \langle\tilde{\Psi}_1(t)|\hat{A}_-(z)|\tilde{\Psi}_1(t)\rangle$:

$$A_+(t<0,z) = \alpha\varphi_o^*(\delta k/\sqrt{2\pi})^{1/2}e^{i\phi_1}\exp\{-\tfrac{1}{2}[c(t-t_o)-z]^2\delta k^2\}$$
$$+ \beta\varphi_o^*(\delta k/\sqrt{2\pi})^{1/2}e^{i\phi_2}\exp\{-\tfrac{1}{2}[c(t-t_o-\tau)-z]^2\delta k^2\}, \quad (23a)$$

$$A_-(t>t',z) = \alpha\varphi_o^*(\delta k/\sqrt{2\pi})^{1/2}\exp\{i(\phi_1-\chi_{12})\}\exp\{-\tfrac{1}{2}[c(t-t')+z]^2\delta k^2\}$$
$$+ \beta\varphi_o^*(\delta k/\sqrt{2\pi})^{1/2}\exp\{i(\phi_2-\chi_{12})\}\exp\{-\tfrac{1}{2}[c(t-t'+\tau)+z]^2\delta k^2\}. \quad (23b)$$

As seen from comparison of the states in Eq. (8) and Eq. (22) as well as of the field's amplitudes in Eq. (23a) and Eq. (23b), the retrieval two wave packet field is a temporally revised copy of the Data single photon field. In particular the second wave packet of the initial Data field with amplitude $\beta$ and phase $\phi_2$ corresponds to the first wave packet in the retrieval field (see Fig. 2) (for pure single photon state $\varphi_o = 0$ one can similarly analyze the field amplitudes $A_+(t<0,z) = {}_a\langle 1|_{ph}\langle 0|\hat{A}_+(z)|\tilde{\Psi}_1(t)\rangle$ and $A_-(t>t',z) = {}_a\langle 1|_{ph}\langle 0|\hat{A}_-(z)|\tilde{\Psi}_1(t)\rangle$). We note also that the both wave packets of retrieval field $A_-(t,z)$ get the same additional quantum phase shift $-\chi_{12}$ which can be controlled by the phases of the two laser pulses and by the temporal evolution of the atomic system during storage time $t_1 < t < t_2$.

### 3. Quantum manipulation of time-bin qubit states

We now discuss in more detail the storage and retrieval of time-bin qubit states based on the quantum memory analyzed in the previous sections. Time-bins are convenient to code quantum information in single or few photon-pulses in optical fibers. For simplicity we restrict ourselves here to qubits (2-dimensional quantum states, see Eq. (8) in the state (7b)) coded in two time-bins labelled |0> and |1>, where |0> corresponds to a wave packet $\psi^{(0)} = f(t)e^{-i\omega_0 t}$ and

|1> to a second wave packet $\psi^{(1)} = f(t-\tau)e^{-i\omega_0 t}$ propagating with the time delay $\tau$ (see Eq. (23a)), $f(t)$ is an arbitrary slowly varying envelope function, and $e^{-i\omega_0 t}$ is the phase factor corresponding to the carrier frequency. By convention the first (in time) time-bin is associated to the basic quantum ket |0> and the later time-bin to |1>.

A general qubit state in Eqs. (7c) and (8) has the form $|\psi\rangle = r|0\rangle + \sqrt{1-r^2}e^{i\varphi}|1\rangle$. If the state $|\psi\rangle$ is prepared by introducing a delay between two parts of an original single wave packet, then the relative phase satisfies $\varphi = \omega_0 \tau$. However, using amplitude and phase modulators in combination with a cw laser beam to generate the time-bin states, one can in principle control the delay $\tau$ and the relative phase $\varphi$ independently. The intensity modulator allows one to cut in the continuous wave envelopes of arbitrary shape, while the phase modulator allows one to change the phase of the continuous wave in-between successive pulses. This way of preparing time-bins makes it clear that the delay $\tau$ and the relative phase $\varphi$ are indeed independent parameters. The first parameter is part of the definition of the basic states upon which the qubits are defined, the s

In accordance with Eq. (23b) after storage in the quantum memory, the wave packet $\psi^{(0)}$ will be transformed into $f(-t)e^{-i\omega_0 t}$, and the wave packet $\psi^{(1)}$ into $f(-t-\tau)e^{-i\omega_0 t}$. The envelope functions are inverted in time while the carrier wave phase factors continue to evolve unchanged. This is a consequence of the fact that the photon is stored as a coherent excitation of the medium without changing its energy. Therefore the phase continues to evolve with the same optical frequency for the whole time that the photon is stored in the memory. The photon is stopped, but the phase of the corresponding atomic excitations continues to "keep the time". Since the time order of the reemitted wave packets is changed, their labels 0 and 1 are exchanged according to the above mentioned convention. However, if the input envelope function $f(t-\tau)$ was associated with a phase factor $e^{i\varphi}$ (corresponding to a carrier wave that is shifted by $\varphi$), then the corresponding output envelope $f(-t-\tau)$ will still be associated with this phase factor. The overall quantum state after the memory is therefore given by $r|1\rangle + \sqrt{1-r^2}e^{i\varphi}|0\rangle$.

To conclude this section let us apply the above discussion to the example illustrated in Fig. 3. A short pulse arrives from the left, passes through an imbalanced interferometer with path length difference $\Delta L$ and the two time-bins are stored in a quantum memory of the type considered in this article. At a later time the time-bins are released, they propagate now from right to left and pass a second time through the same imbalanced interferometer. At the left we have 3 pulses, we shall concentrate on the central one, as only this one exhibits interesting interferences. But now, notice that the pulse that propagates through the short arm on its way to the QM (i.e. the early incoming time-bin, denoted s in Fig. 1) remains stored longer than the pulse that propagates through the long arm (l in Fig. 1), because the early incoming time-bin corresponds to the delayed outcoming time-bin. Consequently, the 3 pulses propagating back toward the left correspond to the following paths in the interferometer: "long-short" (ls), "long-long & short-short" (ll and ss) and "short-long" (sl). The counter-intuitive conclusion is that one contribution to the central pulse passes twice through the long arm, hence acquires the phase $2\alpha = 2\omega_0 \Delta L / c$, while the other contribution passes twice through the short arm!

This setup makes it possible to measure the path length difference $\Delta L$ using pulses much shorter than $\Delta L/c$. In order to achieve this task with a sub-wavelength resolution one usually uses a cw laser with a coherence length longer than $\Delta L$. At first sight it may seem impossible to achieve the same measurement using laser pulses shorter than $\Delta L/c$. But this is not so if one uses, for example, our quantum memory. Indeed, as explained in the previous paragraph, the central pulse coming out of the interferometer is modulated with a phase $2\alpha$, providing thus a direct measurement of the path difference, with even a twofold increased resolution. Where is the necessary coherence of our measuring device? Clearly, it is still necessary to have a reference length or clock whose stability is larger than the measured quantity. However, the usual "cw laser clock" is replaced by the ions in the QM: the QM coherence time must be longer than the path length difference of the investigated interferometer divided by $c$.

### 4. Conclusion

Summarizing we note that the analytical solution in Eq. (21) generalizes the obtained early results in [6, 9, 23] opening thereby the possibility to work with arbitrary quantum states of the Data light fields. Similarly to Eq. (23) one can find that the state of the retrieved field in Eq. (21) has a reversed time evolution in comparison with the initial state in Eq. (7b) while preserving all the amplitude and phase relations between the wave packets in the initial state. We have discussed the storage and retrieval of time-bin qubits in the described quantum memory. We have also shown how the quantum memory can be used for a measurement of the difference of path lengths in an imbalanced interferometer using pulses much shorter than the time delay corresponding to the path-length difference.

We note that the perfect temporal inversion of the Data fields opens a new resource for the quantum manipulations of single photon time-bin qubits which are attractive for applications in quantum optics and quantum communications. The control of the global quantum phase $-\chi_{12}$ of the retrieval field could also be an interesting sensitive tool for local field measurements based on the long-lived dynamics of stored quantum states, which will be studied elsewhere.

Recently [24] a feasible method has been proposed to implement conditionally the QM operation with the fidelity of almost unity even for not so strong atom-photon couplings in the cavity quantum electrodynamics (QED) with one Λ-type atom. Analysis of the QM based on photon echo with CRIB in the QED cavity [25] has shown a considerable decrease of the minimal temporal scale for the QM processes and of the atom-photon coupling due to the enhancement of the photon-atom interaction by the factor $\sim N^{1/2}$ (N is a number of atoms) and due to fast dephasing of the inhomogeneously broadened atomic transition. The QM based on photon echo with CRIB in the QED cavity can be used both for faster and long-lived storage of the quantum light field and for control of the coherent interactions between photons and atomic ensembles. Finally we point out recent experiments [26, 27] that demonstrated the suitability of Er3+-doped proton-exchanged LiNbO3 crystalline waveguides and Er3+-doped silicate fibers for the QM protocol based on photon echo with CRIB.

The work was supported by the grant of Russian Foundation of Basic Research No. 06-02-16822 and by the European Integrated Project QAP (IST-015848).

**Figure's Captions**

Fig.1.
a) light field's frequencies and atomic transitions; b) and c) are the spatial schemes of the light fields for two fixed times. $E_{Data}$ is a Data light field, $E_1$ and $E_2$ are the two control laser fields, $E_{echo}$ is a photon echo field.

Fig.2.
Temporal behavior of the frequency detuning $\Delta_j(t)$ for j-th and j'-th atoms. Bottom insert is given for comparison of the temporal behavior with the interaction of atoms and all the light fields. The amplitudes and phases of the single photon two wave packet fields.
$t<0$): $\alpha$, $\phi_1$ and $\beta$, $\phi_2$ are the amplitudes and phases of first and second wave packets of the initial Data field.
$t \geq t' = t_3 + t_1 - t_o$): $\beta$, $\phi_2' = \phi_2 - \chi_{12}$ and $\alpha$, $\phi_1' = \phi_1 - \chi_{12}$ are the amplitudes and phases of first and second wave packets of the retrieval field, $t_2 = 2t_{inv} - t_1$, $\chi_{12} = \xi_1 - \xi_2 - \omega_{32} t_{21}$.

Fig. 3. This setup allows to measure the path length difference $\Delta L$ using pulses of duration much shorter than $\Delta L/c$. The lines represent a Mach-Zehnder interferometer made out of optical fibers and two 50-50% fiber-couplers; QM=Quantum memory.

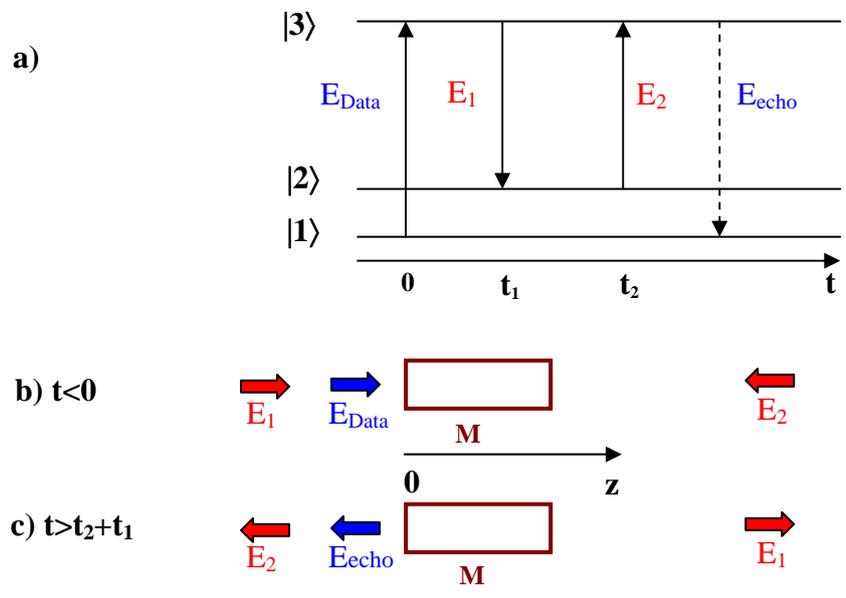

Figure 1.

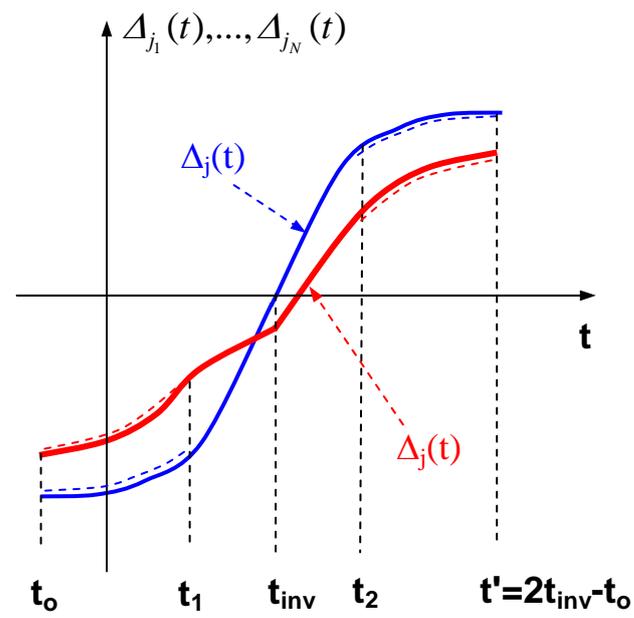

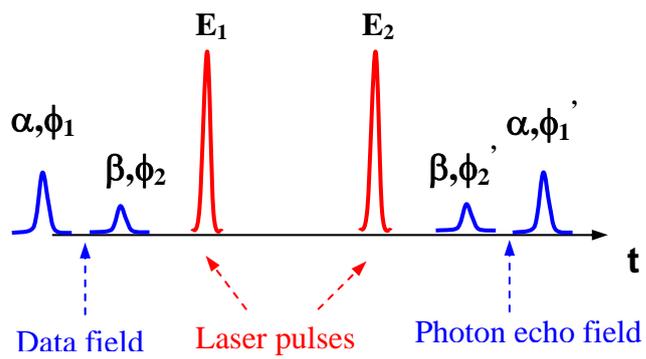

Figure 2.

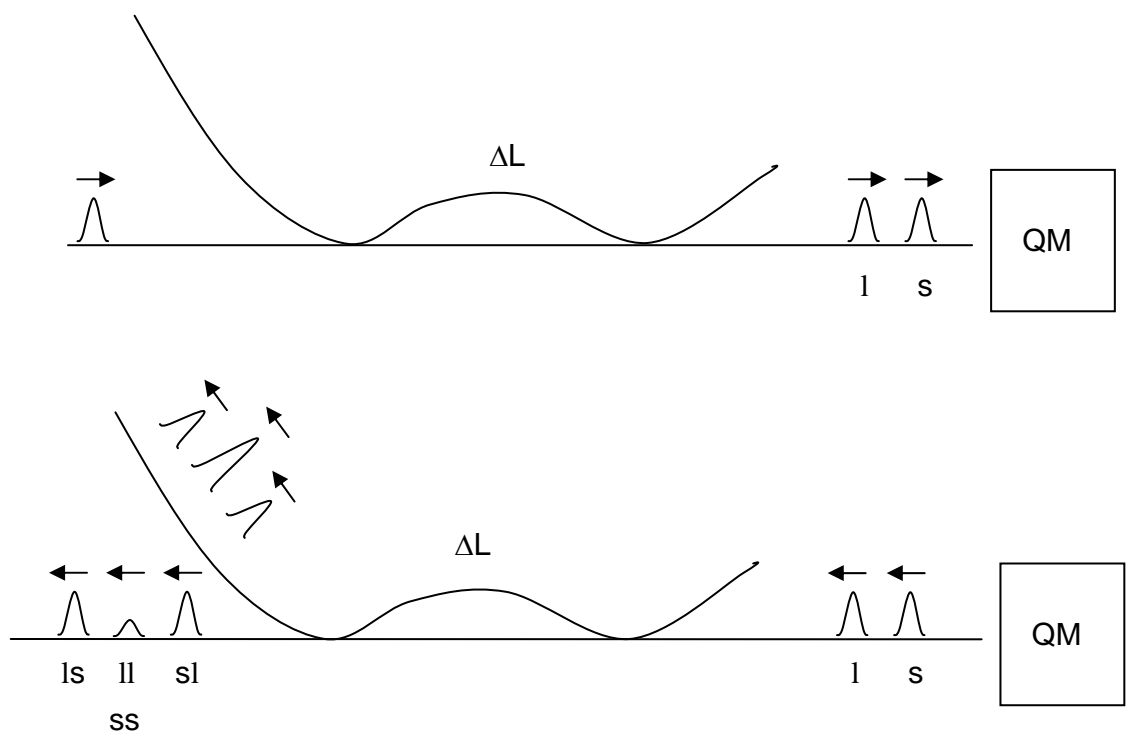

Figure 3.